\documentclass{amsart}

\usepackage{graphicx}

\usepackage[cmtip,all]{xy}

\usepackage[pdftex,linktoc=page]{hyperref} 
\hypersetup{pdfstartview=FitH, pdfpagemode=UseNone}

\def\supp{\text{Supp}}
\def\sym{\text{Symm}}
\def\C{\mathbb{C}}
\def\F{\mathbb{F}}
\def\N{\mathbb{N}}
\def\Z{\mathbb{Z}}
\def\Amc{\mathcal{A}}
\def\Pmc{\mathcal{P}}
\def\Smc{\mathcal{S}}

\DeclareMathOperator{\poly}{poly}

\newtheorem{theorem}{Theorem}[section]
\newtheorem{lemma}[theorem]{Lemma}
\newtheorem{corollary}[theorem]{Corollary}
\newtheorem{conjecture}[theorem]{Conjecture}

\theoremstyle{definition}

\theoremstyle{remark}

\numberwithin{equation}{section}

\newcommand{\comment}[1]{}

\begin{document}

\title[Factoring Using Schemes]{Deterministic Polynomial Factoring and Association Schemes}



\author[Arora]{Manuel Arora}
\address{Hausdorff Center for Mathematics, University of Bonn,
53115 Bonn.}
\email{m.arora@hcm.uni-bonn.de}


\author[Ivanyos]{G\'abor Ivanyos}
\address{Computer and Automation Research Institute
of the Hungarian Academy of Sciences (MTA SZTAKI),
Kende u. 13-17,
H-1111 Budapest, Hungary. }
\email{Gabor.Ivanyos@sztaki.mta.hu}


\author[Karpinski]{Marek Karpinski}
\address{Department of Computer Science, 
University of Bonn, 53117 Bonn.}
\email{marek@cs.uni-bonn.de}


\author[Saxena]{Nitin Saxena}
\address{Hausdorff Center for Mathematics, University of Bonn,
53115 Bonn.}
\email{ns@hcm.uni-bonn.de}

\subjclass[2000]{12Y05, 05E30, 05E10, 03D15, 68W30}

\keywords{algebra decomposition, association scheme, cyclotomic scheme, finite field, GRH, Linnik, 
matching, polynomial factoring, representation theory, smooth number, tensor}

\date{}

\dedicatory{}

\begin{abstract}
The problem of finding a nontrivial factor of a polynomial $f(x)$ over a finite field $\F_q$ has many known efficient, but randomized, 
algorithms. The deterministic complexity of this problem is a famous open question even assuming the generalized 
Riemann hypothesis (GRH). In this work we improve the state of the art by focusing on prime degree polynomials; let $n$ be
the degree. If $(n-1)$ has a `large' $r$-smooth divisor $s$, then we find a nontrivial factor of $f(x)$ in deterministic $\poly(n^r,\log q)$
time; assuming GRH and that $s=\Omega(\sqrt{n/2^r})$. Thus, for $r=O(1)$ our algorithm is polynomial time. Further, for 
$r=\Omega(\log\log n)$ there are infinitely many prime degrees $n$ for which our algorithm is applicable and better than the best known; 
assuming GRH. 

Our methods build on the algebraic-combinatorial framework of $m$-schemes initiated by Ivanyos, Karpinski and Saxena (ISSAC 2009). 
We show that the $m$-scheme on $n$ points, implicitly appearing in our factoring algorithm, has an exceptional structure; leading us
to the improved time complexity. Our structure 
theorem proves the existence of small intersection numbers in any association scheme that has many relations, and roughly equal valencies and indistinguishing numbers. 
\end{abstract}

\maketitle

\tableofcontents


\section{Introduction}

\vspace{-0.02cm}

\addtocontents{toc}{\protect\setcounter{tocdepth}{1}}

We consider the classical problem of finding a nontrivial factor of a given polynomial over a finite field. There exist various randomized polynomial time algorithms for this problem, such as Berlekamp \cite{BE}, Rabin \cite{RA}, Cantor \& Zassenhaus \cite{CZ}, von zur Gathen \& Shoup \cite{VZGS}, Kaltofen \& Shoup \cite{KS}, and Kedlaya \& Umans \cite{KU11}, but its deterministic time complexity is a longstanding open problem. It pertains to the general derandomization question in computational complexity theory, i.e.~whether any problem solvable in probabilistic polynomial time can also be solved in deterministic polynomial time.

In this paper, we consider the deterministic time complexity of the problem of polynomial factoring over finite fields assuming the generalized Riemann hypothesis (GRH) (Section \ref{Algebraic Prerequisites}). GRH enables us to find primitive $r$-th nonresidues in a finite field $\F_q$, which are in turn used to find a root $x$ (if it exists in $\F_q$) of polynomials of the type $x^{r}-a$ over $\F_q$ \cite{AMM}. Assuming GRH, there are many deterministic factoring algorithms known but all of them are super-polynomial time except on special input instances: R\'{o}nyai \cite{RO} showed that under GRH, any polynomial $f(x)\in \Z[x]$ can be factored modulo $p$ deterministically in time polynomial in the order of the Galois group of $f(x)$, except for finitely many primes $p$. R{\'o}nyai's result generalizes previous work by Huang \cite{HUA}, Evdokimov \cite{EV2}, and Adleman, Manders \& Miller \cite{AMM}. Bach, von zur Gathen \& Lenstra \cite{BGL} showed that polynomials over finite fields of characteristic $p$ can be factored in deterministic polynomial time if $\phi_k(p)$ is smooth for some integer $k$, where $\phi_k(p)$ is the $k$-th cyclotomic polynomial. This result generalizes previous work by R{\'o}nyai \cite{RO2}, Mignotte \& Schnorr \cite{MS}, von zur Gathen \cite{VZG}, Camion \cite{CAM}, and Moenck \cite{MOE}.

The line of research which interests us was started by R{\'o}nyai \cite{RO3}. He used GRH to find a nontrivial factor of a polynomial $f(x)\in \F_q[x]$, where $n=\deg f$ has a small prime factor, in deterministic polynomial time. R{\'o}nyai's framework relies on the discovery that finding a nontrivial automorphism in certain algebras (such as $\Amc:=\F_q[x]/f(x)$ and its tensor powers) yields an efficient decomposition of these algebras under GRH. Building on R{\'o}nyai's ideas, Evdokimov \cite{EV} showed that an arbitrary degree $n$ polynomial $f(x)\in \F_q[x]$ can be factored deterministically in time $\poly(\log q, n^{\log n})$ under GRH. This line of approach has since been investigated, in an attempt to either remove GRH \cite{IKRS12} or improve the time complexity, leading to several analytic number theory, algebraic-combinatorial conjectures and special case solutions \cite{CH, GAO, SA, IKS}. 

Our method in this paper, building on \cite{IKS}, encompasses the known algebraic-combinatorial (if not analytic number theory) methods and ends up relating the complexity of polynomial factoring to `purely' combinatorial objects (called \textit{schemes} and {\em intersection numbers}) that are central to the research area of algebraic combinatorics. The methods of \cite{RO3,EV,CH,GAO,SA} arrange the underlying roots of the polynomial in a combinatorial object that satisfies \textit{some} of the defining properties of schemes. This paper contributes to the understanding of schemes by making progress on a related purely combinatorial conjecture, which is naturally connected with polynomial factoring. 

\subsection{Our main result}
We study the problem of finding a nontrivial factor of a polynomial of {\em prime} degree. Intuitively, this case should not be any easier. 
However, it turns out that our combinatorial framework is quite well behaved over prime number of roots and gives an improved time complexity. We call a number $s\in\N$ {\em $r$-smooth} if each prime factor of $s$ is at most $r$.

\begin{theorem}[Factoring]\label{Correctness of the Algorithm Intro}
Let $f(x)$ be a polynomial of prime degree $n$ over $\F_{q}$. Assume $(n-1)$ has a $r$-smooth divisor $s$, with $s\geq \sqrt{n/\ell}+1$ and $\ell\in \N_{>0}$. Then we can find a nontrivial factor of $f(x)$ deterministically in time $\poly(\log q, n^{r+\log\ell})$ under GRH. 
\end{theorem}

Naturally, one asks if there exist infinitely many primes $n$ for which Theorem \ref{Correctness of the Algorithm Intro} is a significant improvement. A well-known number theory conjecture concerning primes in arithmetic progressions is connected to this question (Section \ref{Connection with Linnik's Theorem}). Under the conjecture that $L=2$ is admissible for Linnik's constant \cite{L44},  we prove that there exist infinitely many primes $n$ for which the time complexity in Theorem \ref{Correctness of the Algorithm Intro} is polynomial. Even simply under GRH the factoring algorithm has an improved time complexity over the best known ones, for infinitely many $n$. 

\begin{corollary}[Infinite family]\label{Infinitely Many Primes} Assuming GRH, there exist infinitely many primes $n$ such that every polynomial $f(x)\in \F_q[x]$ of degree $n$ can be factored deterministically in time 
$\poly(\log q, n^{\log\log n})$. 

Further if $L=2$ is admissible for Linnik's constant, then there exist infinitely many primes $n$ such that every polynomial $f(x)\in \F_q[x]$ of degree $n$ can be factored deterministically in time 
$\poly(\log q, n)$.
\end{corollary}

The techniques known before our work do not give a result as strong as ours on this particular infinite family of degrees. The best one could have done before is $\poly(\log q, n^{\log n})$
time, by the general purpose algorithm of Evdokimov \cite{EV}.

\subsection{Idea of $m$-schemes}\label{$m$-Schemes and Polynomial Factoring}

The GRH based algorithm for factoring polynomials over finite fields by Ivanyos, Karpinski and Saxena \cite{IKS} (called \textit{IKS-algorithm} in the following) relies on the use of combinatorial schemes, more specifically $m$-schemes. If we denote $\left[n\right]:=\{1,...,n\}$, then an $m$-scheme can be described as a partition of the set $\left[n\right]^s$, for each $1\leq s\leq m$, which satisfies certain natural properties called compatibility, regularity and invariance (Section \ref{Basic Definitions}). The notion of $m$-scheme is closely related to the concepts of superscheme \cite{SM}, association scheme \cite{BI,ZI1}, coherent configuration \cite{HI}, cellular algebra \cite{WL68} and Krasner algebra \cite{K36}. Curiously, techniques initiated by \cite{WL68} are used in another outstanding problem - deciding graph isomorphism.

The IKS-algorithm (Section \ref{Description of the IKS-Algorithm}) associates to a polynomial $f(x)\in \F_q[x]$ the natural quotient algebra $\Amc:=\F_q[x]/f(x)$ and explicitly calculates special subalgebras of its tensor powers $\Amc^{\otimes s}$ ($1\leq s\leq m$). Through a series of operations on systems of ideals of these algebras (which can be performed efficiently under GRH), the IKS-algorithm either finds a zero divisor in $\Amc$ - which is equivalent to factoring $f(x)$ - or obtains an $m$-scheme from the combinatorial structure of $\Amc^{\otimes s}$ ($1\leq s\leq m$). In the latter case, the $m$-scheme obtained may be interpreted as the `reason' why the IKS-algorithm could not find a zero divisor in $\Amc$. 

It is not difficult to prove that the IKS-algorithm always finds a zero divisor in $\Amc$ if we choose $m$ large enough (viz.~in the range $\log n$), yielding that the IKS-algorithm deterministically factors $f(x)$ in time $\poly(n^{\log n}, \log q)$. Moreover, it is conjectured that even choosing $m$ as constant, say $m=c$ where $c\geq 4$, is enough to find a zero divisor in $\Amc$ (and hence factor $f$), which would give the IKS-algorithm a polynomial running time under GRH. This is the subject of the so-called \textit{schemes conjecture} (Section \ref{The Schemes Conjecture}) on the
existence of {\em matchings} (Sections \ref{Generalized Matchings} \& \ref{From $m$-Schemes to Factoring}).

We remark that the schemes conjecture is a purely algebraic-combinatorial conjecture concerning the structure of certain kinds of $m$-schemes. We also note that the schemes conjecture is already proven for an important class of $m$-schemes, namely the so-called orbit $m$-schemes (Theorem \ref{Schemes Conjecture Orbit Schemes}). In this current work, we prove the schemes conjecture for an interesting class of $m$-schemes on a prime number of points, culminating in a somewhat surprising result about the factorization of prime degree polynomials. Our proof builds on the strong relationship of $m$-schemes and association schemes (Section \ref{$3$-Schemes from Association Schemes}), and involves fundamental structure results about association schemes of prime order by Hanaki \& Uno \cite{HU} and Muzychuk \& Ponomarenko \cite{MP}.

\subsection{Idea of association schemes}\label{Association Schemes}

Underlying Theorem \ref{Correctness of the Algorithm Intro} is a structural result about association schemes with bounded valencies and indistinguishing numbers. Recall \cite{ZI1,MP} that an {\em association scheme} is a pair $(X,G)$ which consists of a finite set $X$ and a partition $G$ of $X\times X$ such that 
\begin{enumerate}
\item 
$G$ contains the {\em identity} relation $1:=\{(x,x) \,|\, x\in X\}$,
\item 
if $g\in G$, then $g^{*}:=\{(y,x) \,|\, (x,y)\in g\}\in G$, and 
\item 
for all $f,g,h\in G$, there exists an {\em intersection number} $c^{h}_{fg}\in \N$ such that for all $(\alpha,\beta)\in h$, $c^{h}_{fg}=\#\{\gamma\in X \,|\, (\alpha,\gamma)\in f,(\gamma,\beta)\in g\}$. 
\end{enumerate}
An element $g\in G$ is called a \textit{relation} (or \textit{color}) of $(X,G)$. We call $\left|X\right|$ the \textit{order} of $(X,G)$. 
For each $g\in G$, we define its {\em valency} $n_g:=c^1_{gg^{*}}$, and its {\em indistinguishing number} $c(g):=\sum_{v\in G} c^{g}_{vv^{*}}$. 

Whenever it helps, an association scheme can also be thought of as a colored directed graph with $X$ as vertices and $G$ as edges. But it is richer in algebraic structure than a graph and often evokes the feeling ``group theory without groups'' \cite{BI}.
Below we formulate our main scheme theory result; it essentially proves that a large number of relations means the existence of small intersection numbers (assuming bounded valency and indistinguishing number). It is vaguely related to the structural results in the literature that concern with the so-called {\em Schurity} of schemes \cite{EP00,EP03,EP09,MP}. We are concerned `merely' with two small intersection 
numbers and hence we are able to work with better parameters.
     
\begin{theorem}[Small intersection numbers]\label{Small Intersection Numbers Intro}
Let $(X,G)$ be an association scheme. Assume there exist $c,k,\ell\in \N$ and $0< \delta_1,\delta'_1,\delta'_2\leq 1$ with $1<\ell < (\delta_1^2/\delta'_1)\cdot k$ such that for all $1\neq g\in G$, \[
	\delta_1\cdot k \leq n_g \leq \delta'_1\cdot k \ \text{ and } \ c(g) \leq \delta'_2\cdot c.  
\] 
If $\left|G\right|\geq 2(\delta'_1/\delta_1)^3\delta'_2\cdot \frac{c}{\ell-1} + 2$ then there exist nontrivial relations $u\ne v,w\ne w'\in G$ such that $0<c^{w}_{u^{*}v}\le c^{w'}_{u^{*}v}<\ell$.
\end{theorem}

The above theorem establishes the existence of small intersection numbers in association schemes where both the valencies and indistinguishing numbers of nontrivial relations are confined to a certain range. Interestingly, we give evidence that the result is optimal (Section \ref{Optimality of our Results}). An important example of association schemes of this type are schemes of prime order (Sections \ref{Schemes of Bounded Valency and Indistinguishing Number} \&
\ref{Optimality of our Results}). There the nontrivial relations have equal valency, say $k$ \cite{HU} and equal indistinguishing numbers ($k-1$) \cite{MP}.

\begin{corollary}[Prime scheme]\label{Small Intersection Numbers in Schemes of Prime Order}
Let $(X,G)$ be an association scheme of prime order $n=\left|X\right|$ and valency $k$. Let $\ell\in \N_{>1}$. If 
$\left|G\right|\geq \frac{2(k-1)}{\ell-1}\,+\,2$ then there exist nontrivial relations $u\ne v,w\ne w'\in G$ such that $0<c^{w}_{u^{*}v} \le c^{w'}_{u^{*}v}<\ell$.  
\end{corollary}

Drawing on the connection of association schemes and $m$-schemes, we deduce from Corollary \ref{Small Intersection Numbers in Schemes of Prime Order} the existence of {\em matchings} in certain $m$-schemes on a prime number of points that helps in algebra decomposition (Section \ref{Factorization of Prime-Degree Polynomials}). This is the prime source of our results in the domain of polynomial factoring.

\subsection{Organization}\label{Organization}

\S \ref{$m$-Schemes} provides an introduction to the notion of $m$-schemes and surveys important results and concepts associated therewith. We put a special emphasis on explaining the connection between association schemes and $m$-schemes (\S \ref{$3$-Schemes from Association Schemes}). In \S \ref{The IKS-algorithm} we describe the IKS-algorithm for factoring polynomials over finite fields, which builds on the theory of $m$-schemes. Theorem \ref{Matching Refinement} delineates how to factor polynomials by exploiting $m$-scheme structure. In \S \ref{Factoring Polynomials of Prime Degree} we prove our main results: Theorem \ref{Correctness of the Algorithm Intro} on the factorization of polynomials of prime degree and Theorem \ref{Small Intersection Numbers Intro} on the existence of small intersection numbers in association schemes with bounded valencies and indistinguishing numbers. In addition, \S \ref{sec-nmbr-thry} explains how Theorem \ref{Correctness of the Algorithm Intro} ties in with the density of primes in arithmetic progressions (\S \ref{Connection with Linnik's Theorem}) and discusses in which sense the bounds given in Theorem \ref{Small Intersection Numbers Intro} are optimal (\S \ref{Optimality of our Results}).

\section{Preliminaries: $m$-schemes}\label{$m$-Schemes}

In this section we define special partitions of the set $\left[n\right]^m$ that we call $m$-schemes on $n$ points. These combinatorial objects were first defined in \cite{IKS}. They occur naturally as part of the IKS-algorithm for factoring polynomials over finite fields. In the following, we give an overview of the basic theory of $m$-schemes.

\subsection{Basic definitions}\label{Basic Definitions}

In this section, we introduce the necessary definitions for the study of $m$-schemes. For reference purposes, the terminology used here is the same as in the paper \cite{IKS}.

\textbf{$s$-tuples:} Throughout this section, $V$ is an arbitrary set of $n$ distinct elements. For $1\leq s\leq n$, we define the set of {\em essential} $s$-tuples by \[
    V^{(s)}:=\{(v_{1},v_{2},\ldots,v_{s}) \,|\, \ v_{1},v_{2},\ldots,v_{s} \ \text{are $s$ distinct elements of $V$}\}.
   \] 

\textbf{Projections:} For $s>1$, we define $s$ projections $\pi^{s}_{1},\pi^{s}_{2},\ldots,\pi^{s}_{s}:V^{(s)}\longrightarrow V^{(s-1)}$ by \[
    \pi^{s}_{i}:(v_{1},\ldots,v_{i-1},v_{i},v_{i+1},\ldots,v_{s})\longrightarrow (v_{1},\ldots,v_{i-1},v_{i+1},\ldots,v_{s}).
   \] Moreover, for $1\leq i_1<\ldots<i_k\leq s$ we define \[
    \pi^s_{i_1,\ldots,i_k}:V^{(s)}\longrightarrow V^{(s-k)}, \ \ \ \pi^s_{i_1,\ldots,i_k}=\pi^{s-k+1}_{i_1}\circ \ldots \circ \pi^{s}_{i_k}.
   \] 

\textbf{Permutations:} The symmetric group on $s$ elements $\sym_{s}$ acts on $V^{(s)}$ in a natural way by permuting the coordinates of the $s$-tuples. More accurately, the action of $\tau\in \sym_{s}$ on $(v_{1},\ldots,v_{i},\ldots,v_{s})\in V^{(s)}$ is defined as \[
    (v_{1},\ldots,v_{i},\ldots,v_{s})^{\tau}:=(v_{1^{\tau}},\ldots,v_{i^{\tau}},\ldots,v_{s^{\tau}}).
   \] 

\textbf{$m$-Collection:} For $1\leq m\leq n$, an \textit{$m$-collection} on $V$ is a set $\Pi$ of partitions $\Pmc_{1},\Pmc_{2},\ldots,\Pmc_{m}$ of $V^{(1)},V^{(2)},\ldots,V^{(m)}$ respectively. 

\textbf{Colors:} For $1\leq s\leq m$, the equivalence relation on $V^{(s)}$ corresponding to the partition $\Pmc_{s}$ will be denoted by $\equiv_{\Pmc_{s}}$.

\smallskip\noindent
Below, we discuss some natural properties of $m$-collections that are relevant to us. In the following, let $\Pi=\{\Pmc_{1},\Pmc_{2},\ldots,\Pmc_{m}\}$ be an $m$-collection on $V$.

\textbf{P1 (Compatibility)}: We say that $\Pi$ is \textit{compatible} at level $1< s\leq m$, if $\bar{u},\bar{v}\in P \in \Pmc_{s}$ implies that for every $1\leq i\leq s$ there exists $Q \in \Pmc_{s-1}$ such that $\pi^{s}_{i}(\bar{u}),\pi^{s}_{i}(\bar{v})\in Q$. 

In other words, if two tuples (at level $s$) have the same color then for every projection the projected tuples (at level $s-1$) have the same color as well. It follows that for a class $P\in \Pmc_{s}$, the sets $\pi^{s}_{i}(P):=\{\pi^{s}_{i}(\bar{v}) \,|\, \bar{v}\in P\}$, for all $1\leq i\leq s$, are colors in $\Pmc_{s-1}$.    

\textbf{P2 (Regularity)}: We call $\Pi$ \textit{regular} at level $1< s\leq m$, if $\bar{u},\bar{v}\in Q\in\Pmc_{s-1}$ implies that for every $1\leq i\leq s$ and for every $P \in \Pmc_{s}$, \[
    \#\{\bar{u}'\in P \,|\, \pi^{s}_{i}(\bar{u}')=\bar{u}\}=\#\{\bar{v}'\in P \,|\, \pi^{s}_{i}(\bar{v}')=\bar{v}\}.
   \]  
   
\textbf{Fibres:} We call the tuples in $P\,\cap\,(\pi^{s}_{i})^{-1}(\bar{u})$ the \textit{$\pi^{s}_{i}$-fibres of $\bar{u}$ in $P$}. So regularity, in other words, means that the cardinalities of the fibres above a tuple depend only on the color of the tuple.    

\textbf{Subdegree:} The above two properties motivate the definition of the \textit{subdegree of a color $P$ over a color $Q$} as $s(P,Q):=\frac{\left|P\right|}{\left|Q\right|}$, assuming that $\pi^{s}_{i_1,\ldots,i_k}(P)=Q$ for some $1\leq i_1<\ldots<i_k\leq s$ and that $\Pi$ is regular at all levels $2,\ldots,s$.  
   
\textbf{P3 (Invariance)}: We say that $\Pi$ is \textit{invariant} at level $1< s\leq m$, if for every $P \in \Pmc_{s}$ and $\tau\in \sym_{s}$, we have: \[
    P^{\tau}:=\{\bar{v}^{\tau} \,|\, \bar{v}\in P\}\in \Pmc_{s}.
   \] In other words, the partitions $\Pmc_{1},\ldots,\Pmc_{m}$ are invariant under the action of the corresponding symmetric group. 

\textbf{P4 (Homogeneity)}: We say that $\Pi$ is \textit{homogeneous} if $\left|\Pmc_{1}\right|=1$.

\textbf{P5 (Antisymmetry)}: We say that $\Pi$ is \textit{antisymmetric} at level $1< s\leq m$, if for every $P \in \Pmc_{s}$ and $id\neq \tau\in \sym_{s}$, we have $P^{\tau}\neq P$.

\textbf{P6 (Symmetry)}: We say that $\Pi$ is \textit{symmetric} at level $1< s\leq m$, if for every $P \in \Pmc_{s}$ and $\tau\in \sym_{s}$, we have $P^{\tau}= P$. 

Note that an $m$-collection is called compatible, regular, invariant, symmetric, or antisymmetric if it is at every level $1< s\leq m$, compatible, regular, invariant, symmetric, or antisymmetric respectively. 

\textbf{$m$-Scheme:} An $m$-collection is called an {\em $m$-scheme} if it is compatible, regular and invariant.

We start with an easy non-existence lemma for $m$-schemes \cite[Lemma 1]{IKS}. Note that the lemma below puts the main content of \cite{RO3} in a more general framework.
  
\begin{lemma}\label{Non-Existence 1}
Let $r>1$ be a divisor of $n$. Then for $m\geq r$ there does not exist a homogeneous and antisymmetric $m$-scheme on $n$ points.
\proof
For $m\geq r$, clearly every $m$-scheme contains an $r$-scheme (hint: Project the tuples to the first $r$ places). Hence it suffices to prove the above statement for $m=r$. Suppose for the sake of contradiction that there exists a homogeneous and antisymmetric $r$-scheme $\Pi=\{\Pmc_{1},\Pmc_{2},\ldots,\Pmc_{r}\}$ on $V=\{v_{1},v_{2},\ldots,v_{n}\}$. By definition, $\Pmc_{r}$ partitions $n(n-1)\cdots(n-r+1)$ tuples of $V^{(r)}$ into, say, $t_{r}$ colors. By antisymmetricity, every such color $P$ has $r!$ associated colors, namely $\{P^{\tau} \,|\, \tau\in \sym_{r}\}$. Moreover, by homogeneity, the size of every color at level $r$ is divisible by $n$. Hence, $r!n|n(n-1)\cdots(n-r+1)$. But this implies $r!|(n-1)\cdots(n-r+1)$, which contradicts $r|n$. Therefore, $\Pi$ cannot exist. \qed  
\end{lemma}

Below, we describe the relationship between $m$-schemes and association schemes.

\subsection{$3$-schemes from association schemes}\label{$3$-Schemes from Association Schemes}

The notion of $m$-schemes is closely related to the concept of association schemes. Association schemes are standard combinatorial objects for which there exists extensive literature \cite{BN39,BM59,D73,BI,ZI1}. We recall some important identities which involve the valencies of association schemes. Note that the identities given below can all be found in \cite{ZI1}. 

\begin{lemma}\label{Intersection Number Identities}
Let $(X,G)$ be an association scheme and let $d,e,f\in G$. The following holds:
	\begin{enumerate}
		\item $c^{f}_{de}=c^{f^{*}}_{e^{*}d^{*}}$,
		\item $c^{e}_{df}\cdot n_e=c^{d}_{ef^{*}}\cdot n_d$,
		\item $\sum_{g\in G} c^{f}_{ge}=n_{e^{*}}$,
		\item $\sum_{g\in G} c^{g}_{ef}\cdot n_g=n_{e}\cdot n_{f}$.
	\end{enumerate}
\end{lemma}   
   
We now show that the concepts of $3$-scheme and association scheme are essentially equivalent (strictly speaking, former is a refinement of the latter).
The following lemma states that the first two levels of any $3$-scheme constitute an association scheme (up to containment of the identity relation).

\begin{lemma}\label{m-schemes and association schemes 1}
Let $\Pi=\{\Pmc_{1},\Pmc_{2},\Pmc_{3}\}$ be a homogeneous $3$-scheme on the set $V=\{v_{1},v_{2},\ldots,v_{n}\}$. Then $\left(\Pmc_{1},\Pmc_{2}\cup\{1\}\right)$ constitutes an association scheme, where $1=\{(v,v) \,|\, v\in V\}$ denotes the identity relation.
\proof
We prove that for all $P_{i},P_{j},P_{k}\in \Pmc_{2}$, there exists an integer $c^{k}_{ij}$ such that for all $(\alpha,\beta)\in P_{k}$, \[
		c^{k}_{ij}=\#\{\gamma\in V \,|\, (\alpha,\gamma)\in P_{i},(\gamma,\beta)\in P_{j}\}.
	\] The trivial case where at least one of $P_{i},P_{j},P_{k}$ is the identity relation is omitted. By the compatibility and regularity of $\Pi$ at level $3$, there exists a subset $\Smc\subseteq \Pmc_{3}$ such that for all $(\alpha,\beta)\in P_{k}$, the set $\{\gamma\in V \,|\, (\alpha,\gamma)\in P_{i},(\gamma,\beta)\in P_{j}\}$ can be partitioned as \[
		\dot{\cup}_{P\in \Smc}\{\gamma\in V \,|\, (\alpha,\gamma)\in P_{i},(\gamma,\beta)\in P_{j},(\alpha,\gamma,\beta)\in P\}.
	\] By the compatibility of $\Pi$ at level $3$, this partition can simply be written as \[
		\dot{\cup}_{P\in \Smc}\,\{\gamma\in V \,|\, (\alpha,\gamma,\beta)\in P\}.
	\] By the regularity of $\Pi$ at level $3$, the size of each set in the above partition is $\frac{\left|P\right|}{\left|P_{k}\right|}$, which means that \[
		\#\{\gamma\in V \,|\, (\alpha,\gamma)\in P_{i},(\gamma,\beta)\in P_{j}\}=\sum_{P\in\Smc}\frac{\left|P\right|}{\left|P_{k}\right|}.
	\] Since the above equation is independent of the choice of $(\alpha,\beta)\in P_{k}$, it follows that $\left(\Pmc_{1},\Pmc_{2}\cup\{1\}\right)$ is an association scheme. \qed  
\end{lemma}	     

The next lemma states that, in turn, every association scheme also naturally gives rise to a $3$-scheme.

\begin{lemma}\label{m-schemes and association schemes 2}
Let $\left(\Pmc_{1},\Pmc_{2}\right)$ be an association scheme on $V=\{v_{1},v_{2},\ldots,v_{n}\}$. Let $\equiv_{\Pmc_{2}}$ denote the equivalence relation on $V\times V$ corresponding to the partition $\Pmc_{2}$. Let $\Pmc_{3}$ be the partition of $V^{(3)}$ such that for two triples $(u_{1},u_{2},u_{3})$ and $(v_{1},v_{2},v_{3})$, we have $(u_{1},u_{2},u_{3})\equiv_{\Pmc_{3}}(v_{1},v_{2},v_{3})$ if and only if \[ 
(u_{1},u_{2})\equiv_{\Pmc_{2}}(v_{1},v_{2}), \ \ \ (u_{1},u_{3})\equiv_{\Pmc_{2}}(v_{1},v_{3}), \ \ \ (u_{2},u_{3})\equiv_{\Pmc_{2}}(v_{2},v_{3}). \] Then $\{\Pmc_{1},\Pmc_{2}-\{1\},\Pmc_{3}\}$ is a $3$-scheme.
\proof
It is an easy exercise to show that $\{\Pmc_{1},\Pmc_{2}-\{1\},\Pmc_{3}\}$ satisfies compatibility, regularity and invariance. \qed           
\end{lemma}  

\subsection{Generalized matchings}\label{Generalized Matchings}

We now define the notion of \textit{matchings}, certain special colors of $m$-schemes that play an important role in the IKS-factoring algorithm described later. This combinatorial object - matching - provides an algebraic object - ideal automorphism. As before, let $V=\{v_{1},v_{2},\ldots,v_{n}\}$ be a set of $n$ distinct elements and let $\Pi=\{\Pmc_{1},\Pmc_{2},\ldots,\Pmc_{m}\}$ be an $m$-scheme on $V$.

\textbf{Matching:} A color $P\in \Pmc_{s}$ at any level $1<s\leq m$ is called a \textit{matching} if there exists $1\leq i_1<\ldots<i_k\leq s$ and $1\leq j_1<\ldots<j_k\leq s$ with $(i_1,\ldots,i_k)\neq (j_1,\ldots,j_k)$ such that $\pi^{s}_{i_1,\ldots,i_k}(P)=\pi^{s}_{j_1,\ldots,j_k}(P)$ and $\left|\pi^{s}_{i_1,\ldots,i_k}(P)\right|=\left|P\right|$.
  
Note that the paper \cite{IKS} which originally defined the concept of matchings had the restriction that $k=1$. The above definition is broader and constitutes a natural generalization of the previous (limited) notion of matchings. The next theorem gives an important sufficient condition for the existence of matchings in $m$-schemes \cite[Lemma 8]{IKS}.    

\begin{theorem}\label{Matching Lemma alt.}
Let $\Pi=\{\Pmc_{1},\Pmc_{2},\ldots,\Pmc_{m}\}$ be an $m$-scheme on $V=\{v_{1},v_{2},\ldots,v_{n}\}$. Assume $\Pi$ is antisymmetric at level $2$. Moreover, assume there exist colors $P_{t}\in \Pmc_{t}$ and $P_{t-1}:=\pi^{t}_i(P_t)\in \Pmc_{t-1}$ for some $1<t<m$ and $1\leq i\leq t$ such that $1<s(P_{t},P_{t-1})=\frac{\left|P_{t}\right|}{\left|P_{t-1}\right|}\le\ell$ and $m\geq t-1 + \log_2 \ell$, where $\ell\in \N$. Then there exists a matching in $\{\Pmc_{1},\Pmc_{2},\ldots,\Pmc_{m}\}$.  
\proof
Wlog, let us assume that $P_{t-1}=\pi^{t}_t(P_t)\in \Pmc_{t-1}$. We outline an iterative way of finding a matching in $\Pi$. Note that the set \[
    U_{t+1}:=\{\bar{v}\in V^{(t+1)} \,|\, \pi^{t+1}_{t}(\bar{v}),\pi^{t+1}_{t+1}(\bar{v})\in P_{t}\}
    \] is a nonempty union of colors in $\Pmc_{t+1}$. Let $P_{t+1}$ be a color of $\Pmc_{t+1}$ such that $P_{t+1}\subseteq U_{t+1}$. Then by the antisymmetry of $\Pi$ we have \[
   s(P_{t+1},P_{t})=\frac{\left|P_{t+1}\right|}{\left|P_{t}\right|}<\frac{s(P_{t},P_{t-1})}{2}\leq\frac{\ell}{2}.
   \] Evidently, if $s(P_{t+1},P_{t})=1$ then $P_{t+1}$ is a matching. Otherwise, if $s(P_{t+1},P_{t})>1$ we proceed to level $t+2$ and again strictly halve the subdegree (by the same argument as above). This procedure finds a matching in at most $\log_2 \ell$ rounds. \qed                  
\end{theorem}

As a corollary to the above theorem, we have that a homogeneous $m$-scheme on $n$ points which is antisymmetric at level $2$ always has a matching if $m\geq\log_2 n$.    

\begin{corollary}\label{Matching Lemma}
Let $\Pi=\{\Pmc_{1},\Pmc_{2},\ldots,\Pmc_{m}\}$ be a homogeneous $m$-scheme on the set $V=\{v_{1},v_{2},\ldots,v_{n}\}$. Let $\Pi$ be antisymmetric at level $2$. If $m\geq \log_2 n$ then there exists a matching in $\{\Pmc_{1},\Pmc_{2},\ldots,\Pmc_{m}\}$.    
\end{corollary}

\subsection{The schemes conjecture}\label{The Schemes Conjecture}

In Corollary \ref{Matching Lemma} it was shown that every antisymmetric $m$-scheme on $n$ points (for large enough $m$) contains a matching between levels $1$ and $\log_2 n$. Below, we formulate a conjecture which asserts the existence of a constant $c\geq 4$ that could replace the above $\log_2 n$-bound.

\smallskip\noindent
\textbf{Schemes conjecture.} \textit{There exists a constant $c\geq 4$ such that every homogeneous, antisymmetric $m$-scheme with $m\geq c$ contains a matching.}

\smallskip
In Section \ref{The IKS-algorithm} we recall \cite{IKS} that, under GRH, the correctness of the schemes conjecture implies a deterministic polynomial time algorithm for the factorization of polynomials over finite fields (Theorem \ref{Matching Refinement}). The schemes conjecture is especially motivated by the fact that it is known to be true for an important class of $m$-schemes, called \textit{orbit schemes}. An exact definition of orbit schemes follows. Let $V=\{v_{1},v_{2},\ldots,v_{n}\}$ be a set of $n$ distinct elements and $G\leq \sym_{V}$ a permutation group. Fix $1\leq m\leq n$. For $1\leq s\leq m$, let $\Pmc_{s}$ be the partition on $V^{(s)}$ such that for any two $s$-tuples $(u_{1},u_{2},\ldots,u_{s})$ and $(v_{1},v_{2},\ldots,v_{s})$, we have $(u_{1},u_{2},\ldots,u_{s})\equiv_{\Pmc_{s}}(v_{1},v_{2},\ldots,v_{s})$ if and only if \[ 
    \exists \ \sigma\in G: \ \ \ (\sigma(u_{1}),\sigma(u_{2}),\ldots,\sigma(u_{s}))=(v_{1},v_{2},\ldots,v_{s}). 
   \] 
Then $\{\Pmc_{1},\Pmc_{2},\ldots,\Pmc_{m}\}$ is an $m$-scheme on $V$. We call $m$-schemes which arise in the above-described manner {\em orbit $m$-schemes}. They suggest that
the notion of $m$-schemes generalizes that of finite permutation groups.

\begin{theorem}[Schemes conjecture for orbit $m$-schemes]\label{Schemes Conjecture Orbit Schemes}
For $m\geq4$, every homogeneous, antisymmetric orbit $m$-scheme contains a matching.            
\proof
This is shown in \cite[Section 4.1]{IKS}. \qed
\end{theorem}

\section{Preliminaries: The IKS-algorithm}\label{The IKS-algorithm}

In this section, we discuss the GRH based IKS-algorithm for factoring polynomials over finite fields \cite{IKS}. It fundamentally relies on the theory of $m$-schemes. It was shown in \cite{IKS} that the IKS-algorithm has a deterministic polynomial running-time for factoring polynomials of prime degree $n$, where $(n-1)$ is a {\em constant-smooth} number. In Section \ref{Factoring Polynomials of Prime Degree}, we significantly improve this result to polynomials of prime degree $n$, where $(n-1)$ has a {\em large} constant-smooth factor. This relaxation implies that under a well-known number theory conjecture involving Linnik's constant, there are infinitely many primes $n$ such that any polynomial $f(x)\in \F_q[x]$ of degree $n$ can be factored by the IKS-algorithm in time $\poly(n,\log q)$.

\subsection{Algebraic prerequisites}\label{Algebraic Prerequisites}  

We now discuss algebraic prerequisites for the description of the IKS-algorithm. Below, we recapitulate some of the basic concepts of polynomial factoring over finite fields.

\textbf{Associated quotient algebra $\Amc$:} In order to solve polynomial factoring over finite fields, it is enough to factor polynomials $f(x)$ of degree $n$ over $\F_{q}$ that have $n$ distinct roots $\alpha_{1},\ldots,\alpha_{n}$ in $\F_{q}$ \cite{BE,BE70}. Given a polynomial $f(x)\in \F_q[x]$, for any field extension $k\supseteq\F_{q}$, we have the \textit{associated quotient algebra} \[
	 \Amc:=k[x]/(f(x)).   	
	\] It is isomorphic to the direct product of $n$ fields. In the following, we interpret $\Amc$ as the algebra of all functions \[
	 V :=\{\alpha_{1},\ldots,\alpha_{n}\}\longrightarrow k.   	
	\] 

\textbf{The factors of $f(x)$ appear as zero divisors in $\Amc$:} Assume $y(x)z(x)=0$ for some nonzero polynomials $y(x),z(x)\in\Amc$. Then $f(x)\,|\,y(x)\cdot z(x)$, which implies $\text{gcd}(f(x),z(x))$ factors $f(x)$ nontrivially. Since the $\text{gcd}$ of polynomials can be computed by the Euclidean algorithm in deterministic polynomial time, factoring $f(x)$ is, up to polynomial time reductions, equivalent to finding a zero divisor in $\Amc$.

\textbf{Ideals of $\Amc$ and roots of $f(x)$:} For an ideal $I$ of $\Amc$, we define the \textit{support} of $I$ as \[
	 \supp(I):=V\setminus\{v\in V \,|\, a(v)=0 \ \ \text{for every $a\in I$}\}.   	
	\] Via the support, ideal decompositions of $\Amc$ induce partitions on the set $V$. This is the subject of the following lemma:
	
\begin{lemma}\label{Ideal Decomposition}
If $I_{1},\ldots,I_{t}$ are pairwise orthogonal ideals of $\Amc$ (i.e. $I_{i}I_{j}=0$ for all $i\neq j$) such that $\Amc=I_{1}+\cdots +I_{t}$, then \[
	 V=\supp(I_{1})\sqcup\cdots\sqcup \supp(I_{t}).   	
	\]  
\end{lemma}	

\textbf{Tensor powers of $\Amc$:} For $1\leq m\leq n$, we denote by $\Amc^{\otimes m}$ the $m$-th tensor power of $\Amc$ (as $k$-modules). We may regard $\Amc^{\otimes m}$ as the algebra of all functions from $V^{m}$ to $k$. In this interpretation, the rank one tensor element $h_{1}\otimes\cdots\otimes h_{m}$ corresponds to a function that maps $(v_{1},\ldots,v_{m})\longrightarrow h_{1}(v_{1})\cdots h_{m}(v_{m})$.

\textbf{Essential part of tensor powers:} We define the \textit{essential part} $\Amc^{(m)}$ of $\Amc^{\otimes m}$ to be the (unique) ideal of $\Amc^{\otimes m}$ consisting of the functions which vanish on all the $m$-tuples $(v_{1},\ldots,v_{m})\in V^{m}$ with $v_{i}=v_{j}$ for some $i\neq j$. One may interpret $\Amc^{(m)}$ as the algebra of all functions $V^{(m)}\longrightarrow k$. 

\textbf{Ideals of $\Amc^{(m)}$ and roots of $f(x)$:} As in the case $m=1$, we define the \textit{support} of an ideal $I$ of $\Amc^{(m)}$ as \[
	 \supp(I):=V^{(m)}\setminus\{\bar{v}\in V^{(m)} \,|\, a(\bar{v})=0 \ \ \text{for every $a\in I$}\}.   	
	\] Using this convention, Lemma \ref{Ideal Decomposition} can be generalized as follows:

\begin{lemma}\label{Ideal Decomposition II}
For $s\leq n$, if $I_{s,1},\ldots,I_{s,t_{s}}$ are pairwise orthogonal ideals of $\Amc^{(s)}$ such that $\Amc^{(s)}=I_{s,1}+\cdots +I_{s,t_{s}}$, then \[
	 V^{(s)}=\supp(I_{s,1})\sqcup\cdots\sqcup \supp(I_{s,t_{s}}).   	
	\] 
\end{lemma}	

\textbf{Connection with GRH:} As we already mentioned, the IKS-algorithm relies on the assumption of the generalized Riemann hypothesis (GRH) \cite{Rie,CHO,BCR}. We formally state the hypothesis below. Recall that a \textit{Dirichlet character}, of {\em order} $k\in\N_{>1}$, is defined as a completely multiplicative arithmetic function $\chi:(\Z,+)\longrightarrow (\C,\cdot)$ such that $\chi(n+k)=\chi(n)$ for all $n$, and $\chi(n)=0$ whenever $\text{gcd}(n,k)>1$. Given a Dirichlet character $\chi$, we define the corresponding \textit{Dirichlet L-function} by \[
	 L(\chi,s)=\sum^{\infty}_{n=1}\frac{\chi(n)}{n^{s}}   	
	\] for all complex numbers $s$ with real part $>1$. By analytic continuation, this function can be extended to a meromorphic function defined on all of $\C$. The generalized Riemann hypothesis asserts that, for every Dirichlet character $\chi$, the zeros of $L(\chi,s)$ in the \textit{critical strip} $0<\text{Re}\ s< 1$ all lie on the \textit{critical line} $\text{Re}\ s=1/2$. 
	
Under the assumption of GRH, R{\'o}nyai \cite{RO} showed that the knowledge of any explicit nontrivial automorphism $\sigma\in\text{Aut}(\Amc)$ of $\Amc$ immediately gives us a nontrivial factor of $f(x)$. The latter result is used in the routine of the IKS-algorithm. In \cite{RO}, the ability of computing {\em radicals} ($r$-th roots for prime $r$) in finite fields is used. This can be done assuming GRH by a result of Huang \cite{Hu84}. Thus, GRH `acts' in fact through Huang's result. The motivating case of a prime field and $r=2$ can be easily explained by Ankeny's theorem \cite{Ank52} on the smallest primitive root.

\subsection{Description of the IKS-algorithm}\label{Description of the IKS-Algorithm}

We will now describe the routine of the IKS-algorithm. In the following, let $f(x)\in \F_{q}[x]$ be a polynomial of degree $n$ having $n$ distinct roots $V=\{\alpha_{1},\ldots,\alpha_{n}\}$ in $\F_{q}$. For some field extension $k\supseteq\F_{q}$, let $\Amc:=k[x]/(f(x))$ be the associated quotient algebra. With regards to the algorithm, we assume $\Amc$ is given by structure constants with respect to some basis $b_{1},\ldots,b_{n}$. It was shown in \cite[Lemma 4]{IKS} that we can efficiently compute the essential parts $\Amc^{(s)}$ ($1\leq s\leq n$).

\begin{lemma}\label{Basis Computation}
A basis for $\Amc^{(m)}=(k[X]/(f(X)))^{(m)}$ over $k\supseteq\F_{q}$ can be computed by a deterministic algorithm in time $\poly(\log\left| k\right|, n^{m})$.    
\end{lemma}

We now proceed to give an overview of the routine of the IKS-algorithm. Namely, we describe how an $m$-scheme can be obtained from the ideal decompositions of the essential parts $\Amc^{(s)}$ ($1\leq s\leq n$). For referential purposes, let us quickly recapitulate the algorithmic data:
  
\textbf{Input:} A polynomial $f(x)\in \F_{q}[x]$ of degree $n$ having $n$ distinct roots  $V=\{\alpha_{1},\ldots,\alpha_{n}\}$ in $\F_{q}$.

Also $1<m\leq n$ is given, and we can assume that we have the smallest field extension $k\supseteq\F_{q}$ having $s$-th nonresidues for all $1\leq s\leq m$ (computing $k$ will take $\poly(\log q, m^{m})$ time under GRH).   

\textbf{Output:} A nontrivial factor of $f(x)$ or a homogeneous, antisymmetric $m$-scheme on $V=\{\alpha_{1},\ldots,\alpha_{n}\}$. (In the latter case we get the $m$-scheme only implicitly via a system of ideals of $\Amc^{(m)}$.)

\textbf{Description of the algorithm:} We define $\Amc^{(1)}=\Amc=k[x]/(f(x))$ and compute the essential parts $\Amc^{(s)}$ ($1<s\leq m$) of the tensor powers of $\Amc$ (this takes $\poly(\log q, n^{m})$ time by Lemma \ref{Basis Computation}).  

\textbf{Automorphisms and ideal decompositions of $\Amc^{(s)}$ $(1<s\leq m)$:} Observe that for each $\tau\in \sym_{s}$, the map defined by \[
	 \tau:\Amc^{(s)}\longrightarrow\Amc^{(s)}, \ \ \ (b_{i_{1}}\otimes\cdots\otimes b_{i_{s}})^{\tau}\longrightarrow b_{i_{1^{\tau}}}\otimes\cdots\otimes b_{i_{s^{\tau}}}   	
	\] is an algebra automorphism of $\Amc^{(s)}$. By \cite{RO}, this knowledge of explicit automorphisms of $\Amc^{(s)}$ can be used to efficiently decompose $\Amc^{(s)}$ under GRH: Namely, one can compute mutually orthogonal ideals $I_{s,1},\ldots,I_{s,t_{s}}$ ($t_{s}\geq 2$) of $\Amc^{(s)}$ such that \[
	 \Amc^{(s)}=I_{s,1}+\cdots +I_{s,t_{s}}.     	
	\] By Lemma \ref{Ideal Decomposition II}, the above decomposition of $\Amc^{(s)}$ induces a partition $\Pmc_{s}$ on $V^{(s)}$: \[
	 \Pmc_{s}:\; V^{(s)}=\supp(I_{s,1})\sqcup\cdots\sqcup \supp(I_{s,t_{s}}).   	
	\] Together with $\Pmc_{1}:=\{V\}$ this yields an $m$-collection $\Pi=\{\Pmc_{1},\Pmc_{2},\ldots,\Pmc_{m}\}$ on $V$.

We will now show how to refine the $m$-collection $\Pi$ to an $m$-scheme using algebraic operations on the ideals $I_{s,i}$ of $\Amc^{(s)}$. To do that, we first need a tool to relate lower level ideals $I_{s-1,i}$ to higher level ideals $I_{s,i'}$.

\textbf{Algebra embeddings $\Amc^{(s-1)}\longrightarrow\Amc^{(s)}$:} For each $1<s \leq m$ we have $s$ natural algebra embeddings $\iota^{s}_{1},\ldots,\iota^{s}_{s}:\Amc^{\otimes (s-1)}\longrightarrow\Amc^{\otimes s}$ which map $b_{i_{1}}\otimes\cdots\otimes b_{i_{s-1}}$ to $b_{i_{1}}\otimes\cdots\otimes b_{i_{j-1}}\otimes 1\otimes b_{i_{j}}\otimes\cdots\otimes b_{i_{s-1}}$ respectively (for the $s$ positions of $1$). By restricting $\iota^{s}_{j}$ to $\Amc^{(s-1)}$ and multiplying its image by the identity element of $\Amc^{(s)}$, we obtain $s$ algebra embeddings  $\Amc^{(s-1)}\longrightarrow\Amc^{(s)}$ denoted also by $\iota^{s}_{1},\ldots,\iota^{s}_{s}$. In the following, we interpret $\iota^{s}_{j}(\Amc^{(s-1)})$ as the set of functions $V^{(s)}\longrightarrow k$ which do not depend on the $j$-th coordinate.   

The algorithm is now best described by explaining the five kinds of refinement procedures which {\em implicitly} refine $\Pi$. (Remember we cannot see $V$ but only have access to it
via the ideal $\left\langle f \right\rangle$.) 

\smallskip
\textbf{R1 (Compatibility):} If for any $1< s \leq m$, for any pair of ideals $I_{s-1,i}$ and $I_{s,i'}$ in the decomposition of $\Amc^{(s-1)}$ and $\Amc^{(s)}$ respectively, and for any $j\in\{1,\ldots,s\}$, the ideal $\iota^{s}_{j}(I_{s-1,i})I_{s,i'}$ is neither zero nor $I_{s,i'}$, then we can efficiently compute a subideal of $I_{s,i'}$ and thus, refine $I_{s,i'}$ and the $m$-collection $\Pi$.   

\textit{Note that R1 fails to refine $\Pi$ only when $\Pi$ is a compatible collection.}

\smallskip
\textbf{R2 (Regularity):} If for any $1< s \leq m$, for any pair of ideals $I_{s-1,i}$ and $I_{s,i'}$ in the decomposition of $\Amc^{(s-1)}$ and $\Amc^{(s)}$ respectively, and for any $j\in\{1,\ldots,s\}$, $\iota^{s}_{j}(I_{s-1,i})I_{s,i'}$ is not a free module over $\iota^{s}_{j}(I_{s-1,i})$, then by trying to find a free basis, we can efficiently compute a zero divisor in $I_{s-1,i}$ and thus, refine $I_{s-1,i}$ and the $m$-collection $\Pi$.    

\smallskip
\textit{Note that R2 fails to refine $\Pi$ only when $\Pi$ is a regular collection.}  

\smallskip
\textbf{R3 (Invariance):} If for some $1< s \leq m$ and some $\tau\in \sym_{s}$ the decomposition of $\Amc^{(s)}$ is not $\tau$-invariant, then we can find two ideals $I_{s,i}$ and $I_{s,i'}$ such that $I^{\tau}_{s,i}\cap I_{s,i'}$ is neither zero nor $I_{s,i'}$; hence, we can efficiently refine $I_{s,i'}$ and the $m$-collection $\Pi$.   

\textit{Note that R3 fails to refine $\Pi$ only when $\Pi$ is an invariant collection.}

\smallskip
\textbf{R4 (Homogeneity):} If the algebra $\Amc^{(1)}=\Amc$ is in a known decomposed form, then we can trivially find a nontrivial factor of $f(x)$ from that decomposition.

\textit{Note that R4 fails to refine $\Pi$ only when $\Pi$ is a homogeneous collection.}

\smallskip
\textbf{R5 (Antisymmetry):} If for some $1< s \leq m$, for some ideal $I_{s,i}$ and for some $\tau\in \sym_{s}\setminus \{id\}$, we have $I^{\tau}_{s,i}=I_{s,i}$, then $\tau$ is an algebra automorphism of $I_{s,i}$. By \cite{RO}, this means we can find a subideal of $I_{s,i}$ efficiently under GRH and hence, refine $I_{s,i}$ and the $m$-collection $\Pi$. 

\textit{Note that R5 fails to refine $\Pi$ only when $\Pi$ is an antisymmetric collection.}

\smallskip
\textbf{Summary}: The algorithm executes the ideal operations R1-R5 described above on $\Amc^{(s)}$ ($1\leq s \leq m$) until either we get a nontrivial factor of $f(x)$ or the underlying $m$-collection $\Pi$ becomes a homogeneous, antisymmetric $m$-scheme on $V$. It is routine to verify that the time complexity of the IKS-algorithm is $\poly(\log q,n^{m})$.

\subsection{From $m$-schemes to factoring}\label{From $m$-Schemes to Factoring}

We saw in the last subsection how to either find a nontrivial factor of a given $f(x)$ or construct an $m$-scheme on the $n$ roots of $f(x)$. In the following, we explain how to deal with the ``bad case'', when we get a homogeneous, antisymmetric $m$-scheme instead of a nontrivial factor. We will see how the properties of homogeneous and antisymmetric $m$-schemes can be used to obtain a nontrivial factorization of $f(x)$ even in this case. The next theorem is of crucial importance (it is \cite[Theorem 7]{IKS} extended to our general notion of matchings). 

\begin{theorem}[Matchings refine]\label{Matching Refinement}
Let $f(x)$ be a polynomial of degree $n$ over $\F_{q}$ having $n$ distinct roots $V=\{\alpha_{1},\ldots,\alpha_{n}\}$ in $\F_{q}$. Assuming GRH, we either find a nontrivial factor of $f(x)$ or we construct a homogeneous, antisymmetric $m$-scheme on $V$ having no matchings, deterministically in time $\poly(\log q, n^{m})$.  
\proof
We apply the algorithm from Section \ref{Description of the IKS-Algorithm}, suppose it yields a homogeneous, antisymmetric $m$-scheme $\Pi=\{\Pmc_{1},\Pmc_{2},\ldots,\Pmc_{m}\}$ on $V$. For the sake of contradiction, assume that some color $P\in\Pmc_{s}$ is a matching. Let $1\leq i_1<\ldots<i_k\leq s$ and  $1\leq j_1<\ldots<j_k\leq s$ with $(i_1,\ldots,i_k)\neq (j_1,\ldots,j_k)$ be such that $\pi^{s}_{i_1,\ldots,i_k}(P)=\pi^{s}_{j_1,\ldots,j_k}(P)$ and $\left|\pi^{s}_{i_1,\ldots,i_k}(P)\right|=\left|P\right|$. Then $\pi^{s}_{i_1,\ldots,i_k}(\pi^{s}_{j_1,\ldots,j_k})^{-1}$ is a nontrivial permutation of $\pi^{s}_{i_1,\ldots,i_k}(P)$. For the corresponding orthogonal ideal decompositions of $\Amc^{(1)},\ldots,\Amc^{(m)}$, this means that the embeddings \[
   \iota^s_{i_1,\ldots,i_k}:=\iota^{s}_{i_1}\circ \ldots \circ \iota^{s-k+1}_{i_k}, \ \ \ \iota^s_{j_1,\ldots,j_k}:=\iota^{s}_{j_1}\circ \ldots \circ \iota^{s-k+1}_{j_k}
\] both give isomorphisms $I_{s-k,l'}\longrightarrow I_{s,l}$, where the ideals $I_{s-k,l'}$ and $I_{s,l}$ correspond to $\pi^{s}_{i_1,\ldots,i_k}(P)$ and $P$, respectively. Hence, the map $(\iota^{s}_{i_1,\ldots,i_k})^{-1}\iota^{s}_{j_1,\ldots,j_k}$ is a nontrivial automorphism of $I_{s-k,l'}$. By \cite{RO}, this means we can find a subideal of $I_{s-k,l'}$ efficiently under GRH and thus, refine the $m$-scheme $\Pi$.               
\qed  
\end{theorem}

Combining the above result with Corollary \ref{Matching Lemma}, we conclude that one can completely factor $f(x)$ in time $\poly(\log q, n^{\log n})$ under GRH. This reproves Evdokimov's result \cite{EV}, which is based on a framework less general than that of $m$-schemes described above. Note that any progress towards the schemes conjecture (Section \ref{The Schemes Conjecture}) will directly result in an improvement of the time complexity of the IKS-algorithm. A proof of the schemes conjecture, for parameter $c$, would imply that the total time taken for the factorization of $f(x)$ would improve to $\poly(\log q, n^{c})$. 

In the special case that $f(x)$ is a polynomial of prime degree $n$, where $(n-1)$ satisfies certain divisibility conditions, we study the structure of association schemes of prime order to show that for a `small' $m$ the `bad' case in Theorem \ref{Matching Refinement} never happens. This is discussed in the following section.

\section{Factoring prime degree polynomials}\label{Factoring Polynomials of Prime Degree}

In this section we show that the IKS-algorithm has polynomial running time for the factorization of polynomials $f(x)\in \F_q[x]$ of prime degree $n$, where $(n-1)$ has a large constant-smooth factor. By this we mean a number $s\in \N$ of magnitude $\sqrt{n/\ell}$ such that $s\vert (n-1)$ and all prime factors of $s$ are smaller than $r$. The exact relationship beween $\ell,r$ and the time will appear later. Previously, the IKS-algorithm was only known to have polynomial running time for the factorization of polynomials of prime degree $n$, where $(n-1)$ is constant-smooth \cite{IKS}. Our new results imply that under a well-known number theory conjecture involving Linnik's constant, there are infinitely many primes $n$ such that any polynomial $f(x)\in \F_q[x]$ of degree $n$ can be factored by the IKS-algorithm in time $\poly(\log q,n)$. As a main tool, we employ structural results about association schemes of prime order, most notably \cite{HU,MP}. 

\subsection{Schemes with bounded valencies and indistinguishing numbers}\label{Schemes of Bounded Valency and Indistinguishing Number}

We now prove Theorem \ref{Small Intersection Numbers Intro}, which concerns the existence of small intersection numbers in association schemes (with bounded valencies and indistinguishing numbers) assuming large number of relations. Note that Theorem \ref{Small Intersection Numbers Intro} is the principal scheme theory result underlying our main theorem about the factorization of prime degree polynomials (Theorem \ref{Correctness of the Algorithm Intro}). It is a counting (in two ways) argument on the graph of the scheme. It is elementary assuming the fundamental theorems about schemes, but it yields a new interesting property for this class of schemes.  
\\

\noindent \textit{Proof of Theorem \ref{Small Intersection Numbers Intro}.} Fix a relation $1\neq u\in G$ and a tuple $(\alpha,\beta)\in u$. For all $v\in G\setminus\{1,u\}$, define \[
	S_{v}:=\{(\alpha',\gamma)\in X^2 \,|\, (\alpha',\beta)\in u; \ (\alpha,\gamma)\ne(\alpha',\gamma)\in v\}.
\] The set $S_{v}$ consists of those tuples $(\alpha',\gamma)\in X^2$ which together with $(\alpha,\beta)$ form a 
non-degenerate quadrilateral of the type seen below. 
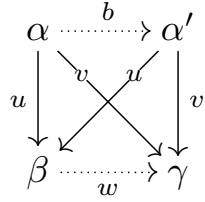
\begin{displaymath}
\scalebox{1.4}{
\xymatrix{
\alpha \ar[d]_u \ar[dr]|<<<<v \ar@{.>}[r]^b & \alpha' \ar[dl]|<<<<u \ar[d]^v \\
\beta \ar@{.>}[r]_w & \gamma
} }
\end{displaymath} 
We determine the cardinality of $S_{v}$. Note that for any relation $b\in G$, there are exactly $c^{u}_{bu}$ choices for $\alpha'\in X$ such that $(\alpha,\alpha')\in b$ and $(\alpha',\beta)\in u$. Moreover, after choosing $\alpha'$, there are exactly $c^{b}_{vv^{*}}$ choices for $\gamma \in X$ such that $(\alpha,\gamma),(\alpha',\gamma)\in v$. Thus, $\left|S_{v}\right|=\sum_{b\in G} c^{u}_{bu} \cdot c^{b}_{vv^{*}}$. Especially, 
\begin{align*}
	\sum_{v\in G\setminus\{1,u\}} \left|S_{v}\right| 
	          = \sum_{1\ne b\in G} c^{u}_{bu} \cdot \sum_{v\in G\setminus\{1,u\}} c^{b}_{vv^{*}}
			\leq \sum_{1\ne b\in G} c^{u}_{bu} \cdot \delta'_2\cdot c
			\leq \delta'_1 \cdot \delta'_2 \cdot c \cdot k,
\end{align*} 
where the last inequality follows from Lemma \ref{Intersection Number Identities} (3).

For the sake of contradiction, assume that for all $v\in G\setminus\{1,u\}$ we have either $c^{w}_{u^{*}v}=0$ or $c^{w}_{u^{*}v}\geq \ell$ for all except at most one relation $w \in G$. We derive a lower bound on $\left|S_{v}\right|$ in order to obtain the contradiction. For $v\in G\setminus\{1,u\}$   define 
$$W_v:=\{w\in G \,|\, c^{w}_{u^{*}v}\neq 0\}.$$ 
Note that for each relation $w\in W_v$ there are exactly $c^{u}_{vw^{*}}$ choices for $\gamma$ such that $(\beta,\gamma)\in w$ and $(\alpha,\gamma)\in v$. Moreover, after choosing $\gamma$, there are exactly $c^{w}_{u^{*}v}-1$ choices for $\alpha'$ such that $(\alpha',\beta)\in u$ and $(\alpha',\gamma)\in v$. Thus, $\left|S_{v}\right|=\sum_{w\in W_v} c^{u}_{vw^{*}}\cdot (c^{w}_{u^{*}v}-1)$. Now observe that $c^{u}_{vw^{*}} \geq c^{w}_{u^{*}v}\cdot \frac{\delta_1}{\delta'_1}$ for all $w\in W_v$ by Lemma \ref{Intersection Number Identities} (1), (2). Since we assume that $c^{w}_{u^{*}v}\geq \ell$ for all except at most one relation $w\in W_v$ we conclude \[
	\left|S_{v}\right|
	   \geq \frac{\delta_1}{\delta'_1} \cdot \sum_{w\in W_v} c^{w}_{u^{*}v}(c^{w}_{u^{*}v}-1) 
	   \geq \frac{\delta_1}{\delta'_1} \cdot \left( (\ell-1)\cdot \sum_{w\in W_v} c^{w}_{u^{*}v} - \frac{\ell^2}{4}\right). 
	\] 
The last inequality is based on the summand-wise inequality: $(\ell-1)c^{w}_{u^{*}v}\, -\,
c^{w}_{u^{*}v}(c^{w}_{u^{*}v}-1)$ $\le (\ell^2/4)$.	
From the equation $\sum_{w\in W_v} c^{w}_{u^{*}v}\cdot n_w=n_{u^{*}} \cdot n_v$ (see Lemma \ref{Intersection Number Identities} (4)) it follows that $\sum_{w\in W_v} c^{w}_{u^{*}v}\geq (\delta_1^2/\delta'_1)\cdot k$. Moreover, using the assumption $1<\ell< (\delta_1^2/\delta'_1)\cdot k$, we deduce 
	\[   
	\left|S_{v}\right|
	   \ge\ \frac{\delta_1}{\delta'_1}\cdot(\ell-1)\cdot \left( \frac{\delta_1^2}{\delta'_1}\cdot k - 
	                 \frac{\ell^2}{4(\ell-1)}\right) 
	      > \frac{\delta_1^3}{2(\delta'_1)^2} \cdot (\ell-1) k.
	\] 
Especially, we have \[
	\sum_{v\in G\setminus\{1,u\}} \left|S_{v}\right|
	                > (\left|G\right|-2)\cdot \frac{\delta_1^3}{2(\delta'_1)^2} \cdot (\ell-1) k.
\] 
This yields $\delta'_1 \delta'_2 \cdot c k > (\left|G\right|-2)\cdot \frac{\delta_1^3}{2(\delta'_1)^2}\cdot(\ell-1)k$ and hence $2(\delta'_1/\delta_1)^3\delta'_2\cdot\frac{c}{\ell-1} + 2 > \left|G\right|$, a contradiction. \qed \\

Let us now consider the special case where $(X,G)$ is an association scheme of prime order $n:=\left|X\right|$. Hanaki-Uno's theorem \cite{HU} tells us that in this case, there exists $k\in \N$ such that $k=n_g$ for all $1\neq g\in G$ (i.e. all nontrivial valencies coincide). We will refer to $k$ simply as the \textit{valency} of $(X,G)$. It was shown in 
\cite[Theorem 3.2]{MP} that for prime order association schemes $(X,G)$ of valency $k$, every nontrivial relation $g\in G$ has indistinguishing number $c(g)=(k-1)$. Combining the above considerations with Theorem \ref{Small Intersection Numbers Intro}, we immediately obtain Corollary \ref{Small Intersection Numbers in Schemes of Prime Order} about prime order association schemes.


\subsection{Factoring algorithm for prime degree polynomials}\label{Factorization of Prime-Degree Polynomials}

Drawing on the scheme theory results from the last subsection, we obtain the following lemma about the existence of matchings in homogeneous antisymmetric $m$-schemes on a prime number of points.

\begin{lemma}\label{Construction of Matching}
Let $\Pi=\{\Pmc_{1},\ldots,\Pmc_{m}\}$ be a homogeneous, antisymmetric $m$-scheme on $V$, where $n:=\left|V\right|$ is a prime number. Let $k$ denote the valency of the association scheme $(\Pmc_{1},\Pmc_{2}\cup\{1\})$. Assume that $m\geq 2\log_2 \ell +3 $ and $\left|\Pmc_{2}\right|\geq \frac{2(k-1)}{\ell-1}+1$ for some $\ell\in \N_{>1}$. Then there exists a matching in $\Pi$.      
\proof
By Corollary \ref{Small Intersection Numbers in Schemes of Prime Order}, there exist nontrivial relations $u\ne v,w\ne w'\in \Pmc_{2}$ such that $0<c^{w}_{u^{*}v}\le c^{w'}_{u^{*}v}<\ell$. Hence there exist $\alpha,\beta,\gamma,\gamma'\in V$ such that $(\alpha,\beta)\in u$, $(\alpha,\gamma),(\alpha,\gamma')\in v$, $(\beta,\gamma)\in w$ and $(\beta,\gamma')\in w'$. Clearly, the relation $P\in\Pmc_{4}$ containing the tuple $(\beta,\alpha,\gamma,\gamma')$ satisfies $\pi^4_{1,3}(P)=\pi^4_{1,4}(P)=v$. Also, $|P|/|v| = |P|/|u| \le c^{w}_{u^{*}v}\cdot c^{w'}_{u^{*}v} \le \ell^2$, thus $P$ has subdegree at most $\ell^2$ over $v$. Now if $s(P,v)=1$ then $P$ is a matching. On the other hand, if $s(P,v)>1$ then we define $Q:=\pi^4_4(P)\in\Pmc_{3}$ and consider the equation $s(P,v)=s(P,Q)\cdot s(Q,v)$. It implies that at least one of the 
subdegrees $s(P,Q), s(Q,v)$ is both at least $2$ and at most $\ell^2$, thus we get a matching in $\Pi$ by suitably invoking Theorem \ref{Matching Lemma alt.}. \qed
\end{lemma}

Using the above lemma about the existence of matchings in $m$-schemes on a prime number of points, we can now prove our main result, Theorem \ref{Correctness of the Algorithm Intro}. \\

\noindent \textit{Proof of Theorem \ref{Correctness of the Algorithm Intro}.} Let $\ell':=(2\ell+1)$.
 It suffices to consider the case that $f(x)$ has $n$ distinct roots $V=\{\alpha_{1},\ldots,\alpha_{n}\}$ in $\F_{q}$. Let $m:=\max\{r+1,2\log_2\ell'+3\}$. We apply the IKS-algorithm (Section \ref{The IKS-algorithm}) and by Theorem \ref{Matching Refinement} either find a nontrivial factor of $f(x)$ or construct a homogeneous, antisymmetric $m$-scheme $\Pi=\{\Pmc_{1},\Pmc_{2},\ldots,\Pmc_{m}\}$ on $V$ having no matchings, deterministically in time $\poly(\log q, n^m)$. Suppose for the sake of contradiction that the latter case occurs. 

Clearly, $(\Pmc_{1},\Pmc_{2}\cup\{1\})$ is an association scheme of prime order $n$, where $1$ denotes the trivial relation. Thus, by Hanaki-Uno's theorem \cite{HU} there exists $k|(n-1)$ such that $\left|P\right|=kn$ for all $P\in\Pmc_{2}$. 
Thus, $|\Pmc_2|=(n-1)/k$. 
We distinguish between the following two cases.

{\bf Case I:} $\gcd(s,k)=1$. Then $|\Pmc_{2}|=(n-1)/k\ge s\ge \sqrt{2n/(\ell'-1)}+1$. Thus, $k<\sqrt{n(\ell'-1)/2}=
\sqrt{2n/(\ell'-1)}\cdot (\ell'-1)/2\le (s-1)(\ell'-1)/2$, implying $|\Pmc_{2}|\ge s>1+\frac{2k}{\ell'-1}$. Especially, $\Pi$ contains a matching by Theorem \ref{Construction of Matching}, contrary to our assumption.

{\bf Case II:} $\gcd(s,k)>1$. The colors in $\{\Pmc_{2},\ldots,\Pmc_{r+1}\}$ can be used to define a homogeneous, antisymmetric $r$-scheme on $k$ points as follows: Pick $P_{0}\in\Pmc_{2}$ and define $V':=\{\alpha\in V \,|\, (\alpha_{1},\alpha)\in P_{0}\}$. Furthermore, define an $r$-collection $\Pi'=\{\Pmc'_{1},\ldots,\Pmc'_{r}\}$ on $V'$ such that for all $1\leq i\leq r$ and for each color $P\in\Pmc_{i+1}$, we put a color $P'\in\Pmc'_{i}$ such that \[
    P':=\{\bar{v}\in V'^{(i)} \,|\, (\alpha_{1},\bar{v})\in P\}.
   \] Then $\left|V'\right|=k$, and $\Pi'=\{\Pmc'_{1},\ldots,\Pmc'_{r}\}$ is a homogeneous, antisymmetric $r$-scheme on $k$ points. On the other hand, by $\gcd(s,k)>1$ we know that $k$ has a prime divisor which is at most $r$; therefore, $\Pi'$ cannot exist by Lemma \ref{Non-Existence 1}. 
\qed \\

We point out in the next section that, under a well-known number theory conjecture involving Linnik's constant, there are infinitely many primes $n$ for which the time complexity in Theorem \ref{Correctness of the Algorithm Intro} is polynomial.

\section{Number theory considerations}\label{sec-nmbr-thry}

\subsection{Primes $n$ of Theorem \ref{Correctness of the Algorithm Intro}}\label{Connection with Linnik's Theorem}

Linnik's theorem in number theory answers a natural question about primes in arithmetic progressions. For coprime integers $a,s$ such that $1\leq a\leq s-1$, let $p(a,s)$ denote the smallest prime in the arithmetic progression $\{a+is\}_i$. Linnik's theorem states that there exist (effective) constants $c,L>0$ such that 
\[	p(a,s)<cs^L.\] 
There has been much effort directed towards determining the smallest admissible value for the \textit{Linnik constant} $L$. The smallest admissible value currently known is $L=5$, as proven by Xylouris \cite{XY}. It has been conjectured numerous times that $L\leq 2$ \cite{SS,KA1,KA2,HB} as noted below.

\begin{conjecture}\label{L=2 Conjecture}
There exists $c>0$ such that for all coprime integers $a,s$ with $1\leq a\leq s-1$, the smallest prime $p(a,s)$ in the arithmetic progression $\{a+is\,|\,i\in\N\}$ satisfies 
$p(a,s)<cs^2$. 
\end{conjecture}

This conjecture is not known to be true under GRH. The result that comes closest to it, is \cite[Theorem 5.3]{BS96}: $p(a,s)<2(s\log s)^2$.

Let us consider how the primes of the type we described in Theorem \ref{Correctness of the Algorithm Intro} relate to $p(1,s)$. This is the subject of Corollary \ref{Infinitely Many Primes}, which we prove below. \\

\noindent \textit{Proof of Corollary \ref{Infinitely Many Primes}.} 
For the first part, we just assume GRH. Let $r\in \N_{>1}$ be a constant and $s\in \N$ a (large enough) $r$-smooth number. 
By \cite[Theorem 5.3]{BS96} there is a prime $n=p(1,s)<2(s\log s)^2$. Thus, $s>\sqrt{n/2}/\log s\ge (\sqrt{n/2}/\log n)+1 = \sqrt{n/(2\log ^2n)}+1$. 
Thus, we can generate infinitely many primes $n$ such that Theorem \ref{Correctness of the Algorithm Intro} applies for $\ell:=\ell(n)=2\log^2n$, and proves 
a time complexity of $\poly(\log q, n^{\log\log n})$.

For the second part, we additionally assume Conjecture \ref{L=2 Conjecture}. Let $r\in \N_{>1}$ be a constant and $s\in \N$ a (large enough) $r$-smooth number.  
By the conjecture there is a prime $n=p(1,s)<cs^2$. Thus, $s>\sqrt{n/c} \ge \sqrt{n/(c+1)}+1$. 
Thus, we can generate infinitely many primes $n$ such that Theorem \ref{Correctness of the Algorithm Intro} applies for $\ell:=(c+1)$, and proves 
a time complexity of $\poly(\log q, n)$. \qed

\subsection{Optimality of Theorem \ref{Small Intersection Numbers Intro}}\label{Optimality of our Results}  

Naturally, one asks if it is possible to further relax the conditions which Theorem \ref{Correctness of the Algorithm Intro} places on the prime number $n$ (i.e. the degree of the polynomial we want to factor). In our current framework, this translates to asking to which extent we can relax the conditions for the existence of small intersection numbers in schemes of bounded valency and indistinguishing number (Theorem \ref{Small Intersection Numbers Intro}). However, the example of the \textit{cyclotomic scheme} below shows that the conditions of Theorem \ref{Small Intersection Numbers Intro} cannot be relaxed (up to constant factors).

Recall the definition of a cyclotomic scheme \cite{D73,GC92}. Let $p$ be a prime and let $e|(p-1)$. Let $\alpha$ be a generator of the multiplicative group $\F_{p}^{*}$  of the field $\F_{p}$. We denote by $\left\langle \alpha^{e} \right\rangle$ the subgroup generated by $\alpha^{e}$. Let $\Pmc:=\{P_{i} \,|\, 0\leq i\leq e\}$ be the partition on $\F_{p}\times \F_{p}$ such that $P_{0}:=\{(x,x) \,|\, x\in \F_{p}\}$ and 
$$P_{i}:=\{(x,y)\in \F_{p}\times \F_{p} \,|\, x-y\in \alpha^{i}\left\langle \alpha^{e} \right\rangle \}$$ 
for $i=1,\ldots,e$. 
Then it can be checked that $(X,G)=(\F_{p},\Pmc)$ is an association scheme. Moreover, the definition of $(\F_{p},\Pmc)$ does not depend on the choice of the generator $\alpha$. We call $(\F_{p},\Pmc)$ the {\em cyclotomic scheme in $(p,e)$}. 

In the following, let $(\F_{p},\Pmc)$ be the cyclotomic scheme in $(p,e)$ as above and let $k:=(p-1)/e$. For nontrivial relations $P_{r},P_{s},P_{t}\in \Pmc$ and $(x,y)\in P_{t}$, we have \begin{align*}
	c_{rs}^{t}&=\#\{z\in \F_{p} \,|\, (x-z)\in \alpha^{r}\left\langle \alpha^{e} \right\rangle, (z-y)\in \alpha^{s}\left\langle \alpha^{e} \right\rangle  \} \\
	&=\#\{(y_1,y_2)\in \F_{p}^*\times \F_{p}^* \,|\, \alpha^{r}y^e_1 + \alpha^{s}y^e_2=(x-y) \} / e^2.
\end{align*} 
We divide by $e^2$ because that is exactly the number of {\em repetitions} of a value $(y_1^e,y_2^e)$ as we vary $y_1,y_2\in\F_p^*$.

By the Hasse-Weil bound \cite{WEI,V05}, we have \[
	\left| \#\{(y_1,y_2)\in \F_{p}\times \F_{p} \,|\, \alpha^{r}y^e_1 + \alpha^{s}y^e_2=(x-y) \} - (p+1) \right|\leq e^2 \sqrt{p}+O(1),
\] from which it follows that \[
	\left| c_{rs}^{t} - \frac{(p+1)}{e^2} \right|\leq \sqrt{p}+ O(1).
\] 
To make the `error' term small, fix $e=k^{1/3}/c \succeq p^{1/4}$ for a (large enough) constant $c\in\N$. 
Now $(p+1)/e^2\ge 2\sqrt{p}$ and we can estimate that $c_{rs}^t > \frac{k}{2e} > (c/2)\cdot k^{2/3} \succeq p^{1/2}$. 
Also, $|G| > e \ge k/(ck^{2/3})$. Thus, we have an association scheme where both the number of relations and the intersection numbers are large, i.e.~ in the range $k^{\frac{1}{3}}$ and $k^{\frac{2}{3}}$, respectively. This matches the parameters
of Corollary \ref{Small Intersection Numbers in Schemes of Prime Order} exactly.

This proves that our scheme theory result, especially Corollary \ref{Small Intersection Numbers in Schemes of Prime Order}, is optimal. But when $\left|G\right|$ is larger than $k^{1/3}$ the Hasse-Weil bound has too large an error. We do not know whether now `small' nonzero intersection numbers start showing up.  

\section{Conclusion}

We studied polynomial factoring over finite fields, under GRH, mainly through algebraic-combinatorial techniques. These are very effective when the polynomial has a prime degree. 
We are able to give an infinite family of prime degrees for which our analysis is much better than the known techniques. 

The main open question here is to extend this study to factor all prime degree polynomials. The key here is to study the underlying $m$-scheme that the factoring algorithm gets `stuck' with.
Its $3$-subscheme is a nice association scheme (it is equivalenced). Since its intersection numbers, and other deeper representation theory invariants, manifest in the higher
levels of the $m$-scheme, the schemes conjecture  (Section \ref{The Schemes Conjecture}) might be approachable. 

Another question is to slightly improve Corollary \ref{Small Intersection Numbers in Schemes of Prime Order}. We do show that it cannot be improved in generality, but that does not rule out the following improvement: There exist at least two 
constant-small intersection numbers when $|G| \approx k/\log k$.
This would be enough to give an infinite family of primes $n$ so that Theorem \ref{Correctness of the Algorithm Intro} has a polynomial time complexity (only assuming GRH).

Finally, we leave the question of extending Theorem \ref{Small Intersection Numbers Intro}, so that it becomes applicable to composite order association schemes, open. Improvements 
there would likely translate to factoring polynomials of new composite degrees.

\section*{Acknowledgements}

We would like to thank Hausdorff Center for Mathematics and the Department of Computer Science, University of Bonn for its support. Especially, for hosting G.I.~for a crucial part of the research, and for
helping organize a related workshop on algebraic-combinatorial techniques.
We thank Sergei Evdokimov, Akihide Hanaki, Mikhail Muzychuk, Ilya Ponomarenko and Paul-Hermann Zieschang for the many fruitful conversations. Especially, M.A.~is grateful
to Ilya for the numerous, still ongoing, discussions, explanations and pointers.


\bibliographystyle{amsalpha}
\bibliography{refs}

\end{document}